\documentclass[times,twocolumn]{aastex6}

\usepackage{color}

\usepackage{etoolbox}
\makeatletter
\patchcmd{\NAT@citex}
  {\@citea\NAT@hyper@{%
     \NAT@nmfmt{\NAT@nm}%
     \hyper@natlinkbreak{\NAT@aysep\NAT@spacechar}{\@citeb\@extra@b@citeb}%
     \NAT@date}}
  {\@citea\NAT@nmfmt{\NAT@nm}%
   \NAT@aysep\NAT@spacechar\NAT@hyper@{\NAT@date}}{}{}
\patchcmd{\NAT@citex}
  {\@citea\NAT@hyper@{%
     \NAT@nmfmt{\NAT@nm}%
     \hyper@natlinkbreak{\NAT@spacechar\NAT@@open\if*#1*\else#1\NAT@spacechar\fi}%
       {\@citeb\@extra@b@citeb}%
     \NAT@date}}
  {\@citea\NAT@nmfmt{\NAT@nm}%
   \NAT@spacechar\NAT@@open\if*#1*\else#1\NAT@spacechar\fi\NAT@hyper@{\NAT@date}}
  {}{}
\makeatother

\shorttitle{Current-Neutralization, Shear, and Eruptive Activity}
\shortauthors{Liu et al.}

\begin{document}

\title{Electric-Current Neutralization, Magnetic Shear, and Eruptive Activity \\in Solar Active Regions}
\author{
Yang Liu\altaffilmark{1},
Xudong Sun\altaffilmark{1}, 
Tibor T\"or\"ok\altaffilmark{2},  
Viacheslav S. Titov\altaffilmark{2},
James E. Leake\altaffilmark{3}}
\affil{
$^1$W. W. Hansen Experimental Physics Laboratory, Stanford University,  Stanford, CA 94305-4085\\
$^2$Predictive Science Inc., 9990 Mesa Rim Rd., Ste. 170, San Diego, CA 92121, USA\\
$^3$NASA Goddard Space Flight Center, Greenbelt, MD 20771
}


\begin{abstract}
The physical conditions that determine whether or not solar active regions (ARs) produce strong flares and coronal mass ejections (CMEs) are not yet well understood. Here we investigate the association between electric-current neutralization, magnetic shear along polarity inversion lines (PILs), and eruptive activity in four ARs; two emerging and two well-developed ones. We find that the CME-producing ARs are characterized by a strongly non-neutralized total current, while the total current in the ARs that did not produce CMEs is almost perfectly neutralized. The difference in the PIL-shear between these two groups is much less pronounced, which suggests that the degree of current-neutralization may serve as a better proxy for assessing the ability of ARs to produce CMEs.   
\end{abstract}

\keywords{Sun: magnetic fields -- Sun: corona -- Sun: flares -- Sun: coronal mass ejections (CMEs)}

\section{Introduction}
\label{s:int}
It is well established that solar flares and coronal mass ejections (CMEs) are powered by the free magnetic energy stored in volumetric electric currents in the corona, predominantly in active regions (ARs). However, it remains unclear how well these currents are \textit{neutralized}. Full neutralization requires the ``direct'' coronal currents that connect the AR polarity centers to be surrounded by ``return'' currents of equal total strength and opposite direction, i.e., the total current has to be zero not only for the whole AR (which is always approximately the case for well-isolated ARs), but also for each polarity individually. \cite{par96b} suggested that ARs are comprised of small, magnetically isolated flux tubes which are individually current-neutralized, so that a whole AR must be neutralized as well. \citet{mel91,mel95}, however, argued that net (non-neutralized) currents can emerge from the solar interior with the emergence of magnetic flux. \cite{longcope00} devised a simplified model of flux-tube emergence that suggests that most return currents are trapped below or at the photosphere, and \cite{wheatland00} found most of the 21 ARs he studied to be non-neutralized; both works support Melrose's scenario. However, even under the assumption that the total current in an AR is organized essentially in the form of a single coronal flux rope, it is still often hypothesized that the current has to be neutralized since it is formed by localized flows and/or the emergence of a neutralized sub-photospheric flux rope \citep[see][]{torok14a}.
 
The distribution of direct and return currents in ARs may be relevant for their ability to produce CMEs \citep[][]{forbes10}. The existence of non-potential magnetic fields prior to CMEs, indicated for instance by magnetic shear along PILs \citep[e.g.,][]{mackay10}, implies the presence of an electric current that is likely concentrated in a relatively narrow channel above the PIL, presumably in form of a flux rope, embedded into an approximately potential magnetic field. The equilibrium and stability properties of such a flux rope depend largely on the ``hoop force'', referred to a unit length of the rope. The hoop force is a ``self-force'' that solely originates from the interaction of the current elements in the rope with one another. In equilibrium, this force is counterbalanced by the additional interaction of the current elements with the ambient magnetic field, which in this case plays the role of a ``strapping'' field. { A first-order term of expansion in $a/R_c$, where $a$ is the cross-sectional radius and $R_c$ the axis-curvature radius of the rope, shows that} the hoop force is proportional to the square of the total current \citep{Zakharov1986}, which, in turn, is an algebraic sum of direct and return currents. Thus, at least in this approximation, the hoop force vanishes for current-neutralized flux ropes. { Under the assumption that ARs have to be current-neutralized, it is therefore sometimes argued that mechanisms that invoke the hoop force, such as the torus instability \citep{kliem06}, cannot play a role in the development of CMEs.}

{ However, in order to alter the hoop force to a point that deviations from equilibrium are inhibited, the return current likely has to be organized in a coherent and compact manner around the direct-current channel \citep[][]{forbes10}. As we shall see, this does not seem to be the case for real ARs.} Moreover, MHD simulations of coronal flux-rope formation by photospheric vortex flows showed that current-neutralization only occurs if the vortices are relatively far away from the PIL. If they are close enough to produce magnetic shear along the PIL, the direct and return currents become imbalanced, and net-currents develop around the PIL \citep{torok03,dalmasse15}. Simulations of the emergence of a current-neutralized flux rope into the corona show a similar effect: significant PIL-shear develops, and the flux-rope-current becomes non-neutralized \citep{torok14a}. In both cases, the hoop force in the coronal flux rope does not vanish and unstable deviations from equilibrium leading to CMEs become possible.   

{ These theoretical considerations suggest a relationship between the degree of current-neutralization, the amount of PIL-shear, and the eruptive activity of ARs, which has not yet been explored systematically using real data.} Observationally, the distribution of direct and return currents can be inferred from photospheric vector magnetograms by calculating the vertical current density. \citet{venkatakrishnan09} and \citet{gosain14} considered isolated sunspots, which were found to be well-neutralized in most cases. { For ARs, neutralized currents were found in one quiet region \citep[NOAA AR 10940:][]{georgoulis12}, and strong net-currents in two highly-eruptive regions \citep[ARs 10930 and 11158:][]{ravindra11,georgoulis12,petrie13,janvier14a,vemareddy15}, which supports the aforementioned simulation results.}

Here we carry out a pilot study to explore the aforementioned relationship. We investigate four ARs, two of which produced CMEs. In order to make our small sample as representative as possible, we consider two emerging and two well-developed ARs. We describe our data processing and the computation of the relevant quantities in Section 2. In Section 3 we present our results, which are summarized and discussed in Section 4. 

\section{Data and Calculations}
\label{s:obs}

\subsection{Data}
We consider a representative sample of two emerging and two well-developed ARs, two of which produced CMEs. The emerging regions are AR 11158 and AR 11072; the developed ones AR 11429 and AR 12192. AR 11072 is a simple bipolar region that did not produce flares larger than C-class during its disk passage \citep{liu12}. In contrast, the complex ARs 11158, 11429, and 12192 were very active \citep[e.g.,][]{sun12, wan12, ash14, liu14, sun15, tha15}. AR 11158 produced three M3 flares or larger, including one X2.2 flare. AR 11429 produced seven M3 flares or larger, including three X-class flares. All these flares were associated with CMEs. On the other hand, AR 12192 produced seventeen M3 flares or larger during its disk passage, including six X-class flares, but no CMEs were detected to originate from the flare-productive core-region of the AR. 

We use the {\em Solar Dynamics Observatory}'s \citep[SDO;][]{pesnell12} {\it Helioseismic and Magnetic Imager} \citep[HMI;][]{sche12,scho12} vector magnetic field data \citep{hoe14} to calculate the electric-current distribution and the PIL magnetic shear for our ARs. 

\subsection{Calculation of Current-Neutralization}
\label{ss:calc_RCDC}
{ For each AR, the vertical electric-current density, $J_z = \mu_0^{-1} (\frac{\partial By}{\partial x} -  \frac{\partial Bx}{\partial y})$, is derived from the processed HMI vector magnetic field data. The direct, $DC$, and return, $RC$, currents are computed for each polarity by integrating $J_z$ values of different sign separately. In order to associate the correct sign of $J_z$ to the $DC$ (and hence to the $RC$), we search for the dominant sign of $J_z/B_z$ in the polarity centers and in the vicinity of the PIL, above which the direct-current channel is assumed to reside. For ARs 11158 and 11429 the association is very clear, while for ARs 11072 and 12192 it is not. However, as we will show, the quantity relevant for our investigation, $|DC/RC|$, is close to unity for the two latter ARs, making the choice of the sign of $J_z$ interchangeable.} 

Complex ARs typically contain significant non-eruptive flux that is irrelevant for our investigation. Such flux may connect to other polarities of the region, to neighboring ARs, or to remote areas on the Sun. Thus, to select the integration area for complex ARs, we employ a combination of non-linear force-free field (NLFFF) extrapolations \citep{wiegelmann04} and photospheric maps of the squashing factor $Q$ \citep{tit07,tit11}.
 
\begin{figure*}
\centerline{\includegraphics[width=0.74\textwidth]{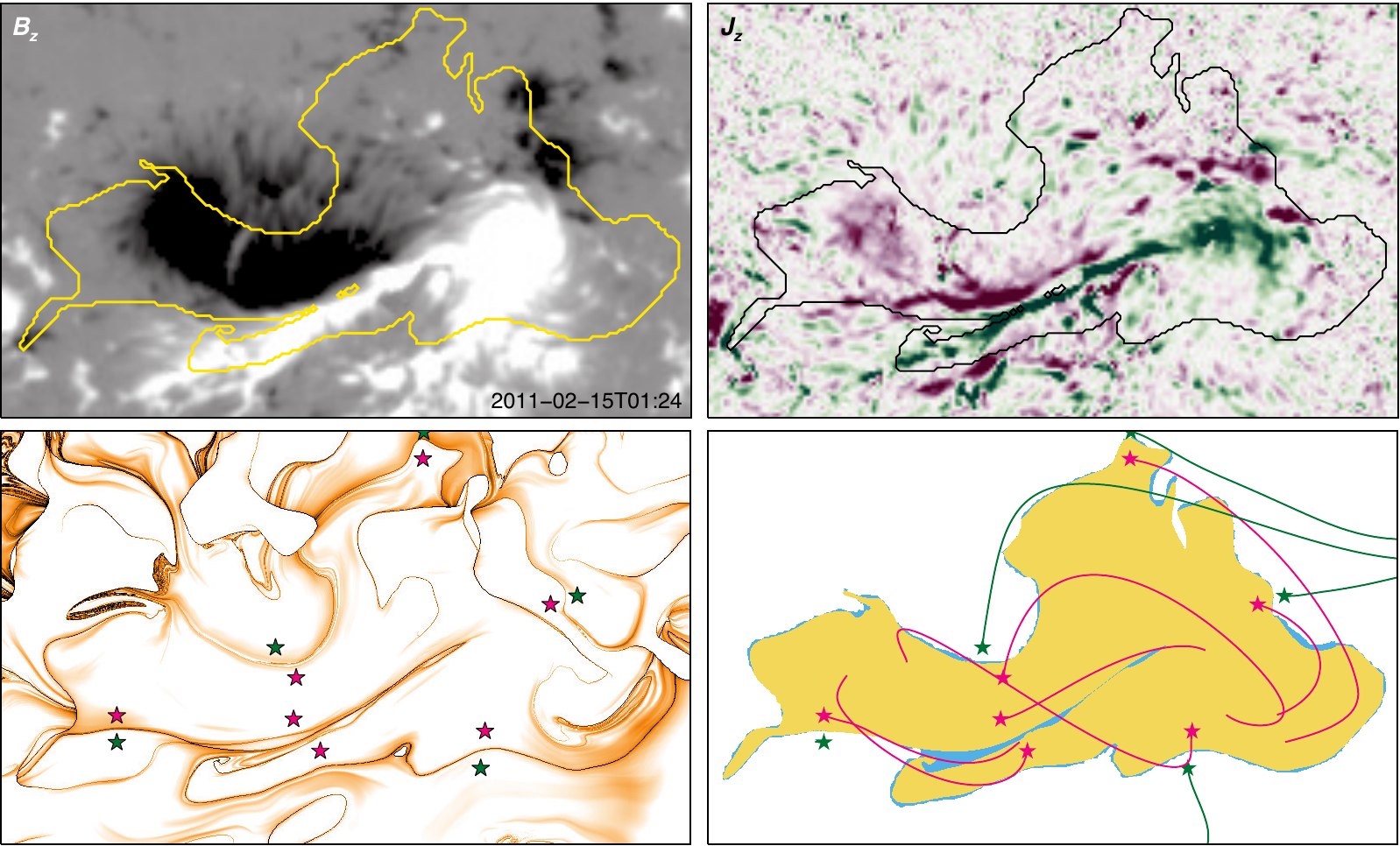}}
\caption
{
{ Top}: Maps of $B_z$ ({\it left}) and $J_z$ ({\it right}) for AR 11158 shortly
before the X2.2 flare, with the $J_z$-integration area shown as contours. $B_z$ is 
scaled to $\pm$ 1000 G; $J_z$ to $\pm 44$ mA $cm^{-2}$. Green (purple) 
colors show positive (negative) values of $J_z$. 
{ Bottom}: Identification of the integration area.
{\it Left:} Q-map at the lower boundary based on an NLFFF model. Dark lanes are
high-Q contours. Pairs of stars indicate magnetic-connectivity boundaries.
Pink (green) stars are field-line foot-points that close within the core region
(connect to the outer AR polarities or remote areas).
{\it Right:} The yellow region shows the final area; cyan regions show 
pixels that were removed from the initially extracted region. Selected field lines
are shown to illustrate the connectivities.
}
\label{magcurrent}
\end{figure*}

As an example, Figure\,\ref{magcurrent} shows the obtained integration area for AR 11158. Only the inner polarity pair of this quadrupolar AR is shown, as the outer polarities were not involved in the region's eruptive activity \citep{sun12}. We first calculate $Q$ at the lower boundary using an NLFFF model. While the pattern of $Q$ is complicated, one can identify an elongated patch encircled by high-$Q$ contours. By tracing field lines along this contour, we see that it outlines the boundary of closed flux connecting the polarities. We then manually extract a slightly larger region and trace field lines from each pixel; those with end points outside of the region are eliminated. Therefore all remaining field lines close within the region. To obtain the final integration area, we apply two common morphology operators, dilation and erosion, to the input image, in order to remove noise, isolate individual elements, and join disparate elements. The final integration area covers most of the polarities, yet excludes significant flux, especially in the positive polarity. The resulting $|DC/RC|$ ratios are \( |DC/RC|^+ = 2.376\pm 0.045\) and $|DC/RC|^- = 3.475\pm 0.099$, respectively. The errors are estimated from the uncertainties of the vector magnetic field data \citep{hoe14}. For comparison, the ratios obtained from the whole area shown in Figure\,\ref{magcurrent} are \( |DC/RC|^+ = 1.997\pm 0.030 \) and $|DC/RC|^- = 2.339\pm 0.046$, which demonstrates that $|DC/RC|$ can significantly increase if irrelevant flux is excluded.

\subsection{Calculation of PIL Magnetic Shear}
In order to estimate the magnetic shear along the main PIL of an AR, we calculate the average magnetic shear angle, $\Phi$, at and around the PIL. To this end, we first locate the PIL in the integration area for each vector magnetogram. The pixels on the PIL are set to be 1; remaining pixels are set to 0. The PIL is then expanded by dilating a 3$\times$3 element. The resulting image includes non-zero pixels at and around the PIL, and zero pixels everywhere else. We then use this mask to compute $\Phi$, by averaging the shear over all non-zero pixels where the field strength is larger than 300 G, which is 3$\sigma$ of the vector magnetic field data noise \citep{hoe14}. The shear for each pixel is calculated following previous studies \citep{hag84,wan94}, which defined it by the angle between the horizontal components of the observed field and a potential field that is computed from the observed vertical field using Fourier transforms \citep{alissandrakis81,gary89}.

\section{Results}
\label{s:res}

\subsection{Two Emerging ARs}
\label{ss:AR_emerge}
We first analyze the two emerging ARs for a time period of six days, which covers the entire emerging process and some evolution afterwards. Figure\,\ref{reversal} shows the ARs, together with a flux-emergence simulation \citep{leake13} that includes an existing, arcade-like coronal magnetic field, for comparison. AR 11072 is a simple bipolar region, which emerges quickly within two days and evolves slowly afterwards. $|DC/RC|$ remains close to unity during the evolution, i.e., the currents remain almost perfectly neutralized. No clear large-scale, coherent current pattern develops. Rather, the direct and return currents are mostly distributed in small, alternating patches \citep[see also][]{cheng.x16}. In contrast, AR 11158 has a complex quadrupolar magnetic configuration. The region is still emerging at the end of our 6-day observation period. As mentioned above, only the inner polarity pair is relevant here. After the onset of flux emergence, $|DC/RC|$ becomes significantly larger than unity, i.e., the AR becomes strongly non-neutralized \citep[cf.][]{janvier14a}. As for AR 11072, patches of alternating direct and return current develop. However, now also an organized, double-J-shaped pattern of strong direct currents forms around the PIL, in very good agreement with the MHD simulation. Analyzing this simulation, \cite{torok14a} found that, as the rising flux temporarily halts before it breaches the photosphere, most of the return current is pushed aside by the subjacent direct current and remains trapped below the surface. 

We use two datasets per day, 12 hours apart, to calculate the magnetic shear for a 5-day time period after the ARs began to emerge. The daily averages are listed in Table\,\ref{tbl-1}, together with the daily averages of $|DC/RC|$. The PIL-shear is much stronger in AR 11158 than in AR 11072. The average shear-angle is $\Phi=63.6^\circ \pm 6.9^\circ$ for AR 11158, compared to only $\Phi=25.1^\circ \pm 6.4^\circ$ for AR 11072.

\subsection{Two Well-Developed ARs}
Figure\,\ref{twoars} shows  the two developed ARs. As before, we analyze the evolution for a 5-day period, but use only one dataset per day because, unlike emerging ARs, developed ARs evolve much more slowly and one dataset per day is sufficient for the purpose of this study. Both ARs were near the central meridian on the third day of this period ($N17S03$ at 00:00 UT on 9 March 2012 and $S12W01$ at 16:00 UT on 23 October 2014 for ARs 11429 and 12192, respectively), and within $35^\circ$ from the central meridian on all other days. The yellow (black) dashed line contours show the integration areas within which $DC/RC$ is calculated. Although both ARs have complex magnetic configurations, $|DC/RC|$ is significantly larger than unity for 11429 during the whole time period, while it remains close to unity for AR 12192 at all times. As for the emerging ARs described in \S\,\ref{ss:AR_emerge}, the current-distribution is characterized by small alternating patches of direct and return currents, and the current-neutralized region (AR 12192) does not show a coherent, large-scale pattern of direct current around its PIL, while the non-neutralized region (AR 11429) does (albeit not as pronounced as in AR 11158). 

Our measurements of the PIL-shear and $|DC/RC|$ are shown in Table\,\ref{tbl-1}. The average shear angle of AR 11429 is $\Phi=67.4^\circ \pm 8.5^\circ$, almost identical to the value obtained for the likewise CME-producing AR 11158. For AR 12192, we find an average shear angle of $\Phi=41.2^\circ \pm 11.3^\circ$, which is significantly larger than for the likewise current-neutralized AR 11072. The 5-day-averaged values of $|DC/RC|$ for the two regions are $2.17\pm 0.01$ and $1.06\pm 0.01$, respectively.

\section{Discussion}
We investigated the possible existence of a relationship between electric-current neutralization, PIL magnetic shear, and the eruptive activity of solar ARs. To this end, we considered a sample of four ARs, including emerging and well-developed ones, as well as quiet and eruptive ones. Using HMI vector magnetic field data, we obtained the electric-current distribution, measured the magnetic shear along the main PIL, and tracked eruptive activity over a time period of 5--6 days for each AR. Table\,\ref{tbl-2} shows an overview of the AR properties relevant for our investigation. Our results can be summarized as follows.

(1) No AR showed signatures of a coherent pattern of return current that would surround a central direct current, as often assumed in theoretical considerations (see the Introduction). Rather, the return currents were distributed in small patches, alternating with neighboring patches of direct current. On the other hand, two ARs had an organized pattern of direct current, which manifests itself as an elongated, double-J-shaped structure bracketing the PIL.

(2) The most striking differences can be found between the ARs that produce CMEs and those that do not. The two CME-producing ARs (one emerging, one developed) are characterized by $|DC/RC|$ values clearly exceeding unity\footnote{{Recently \cite{cheng.x16} reported significantly smaller $|DC/RC|$ values (1.0--1.4) for four CME-producing ARs, but they considered only one AR polarity and did not exclude irrelevant flux (see our Section\,\ref{ss:calc_RCDC}). As expected, $|DC/RC|$ increased to 1.2--2.0 when they restricted the integration to roughly estimated foot-points areas of the eruptions.}}, the presence of a double-J-shaped direct-current channel around the PIL, and an almost identical PIL-shear of about $65^\circ$. In contrast, the ARs that did not produce CMEs (one emerging, one developed) were characterized by $|DC/RC| \approx 1$, i.e., almost perfect current-neutralization, the absence of a direct-current channel around the PIL, and significantly smaller PIL-shear. Out of the quantities we considered, these two ARs differed only in their PIL-shear, which was about $25^\circ$ for the AR that produced no significant flares, and about $40^\circ$ for the one that did. 
        
{ Our results support the association between strong PIL-shear and substantial net-currents found in simulations (Section\,\ref{s:int}). This association is seemingly not simple, though, as considerable shear was present also in our current-neutralized ARs. Furthermore, they suggest that return currents are not relevant for an AR's ability to produce CMEs, as those currents do not form coherent structures that could substantially alter the hoop force in a direct-current channel above the AR's eruptive PIL. The reason why the current-neutralized ARs in our sample do not produce CMEs rather seems to be the absence of such a channel. This supports CME-models that invoke net currents and the hoop force \citep[see the discussion in][]{torok14a}}.

Finally, our results indicate that $|DC/RC|$ may constitute a better quantity for assessing an AR's ability to produce CMEs than the PIL-shear, as the differences in the former quantity appear more pronounced. Equivalently, the presence or absence of a double-J pattern of direct currents around the PIL may indicate whether or not an AR will produce CMEs. On the other hand, it appears that current-neutralized ARs without such a pattern can still produce strong flares, as AR 12192 demonstrates. Those flares may be merely confined, though, and their occurrence may require a certain amount of PIL-shear (the shear in AR 12192 was almost twice as large as in the quiet AR 11072). It has to be kept in mind, though, that very strong flares without CMEs are quite rare, and that AR 12192 was exceptional in this respect \citep[e.g.,][]{sun15}.

\begin{figure*}
\centerline{\includegraphics{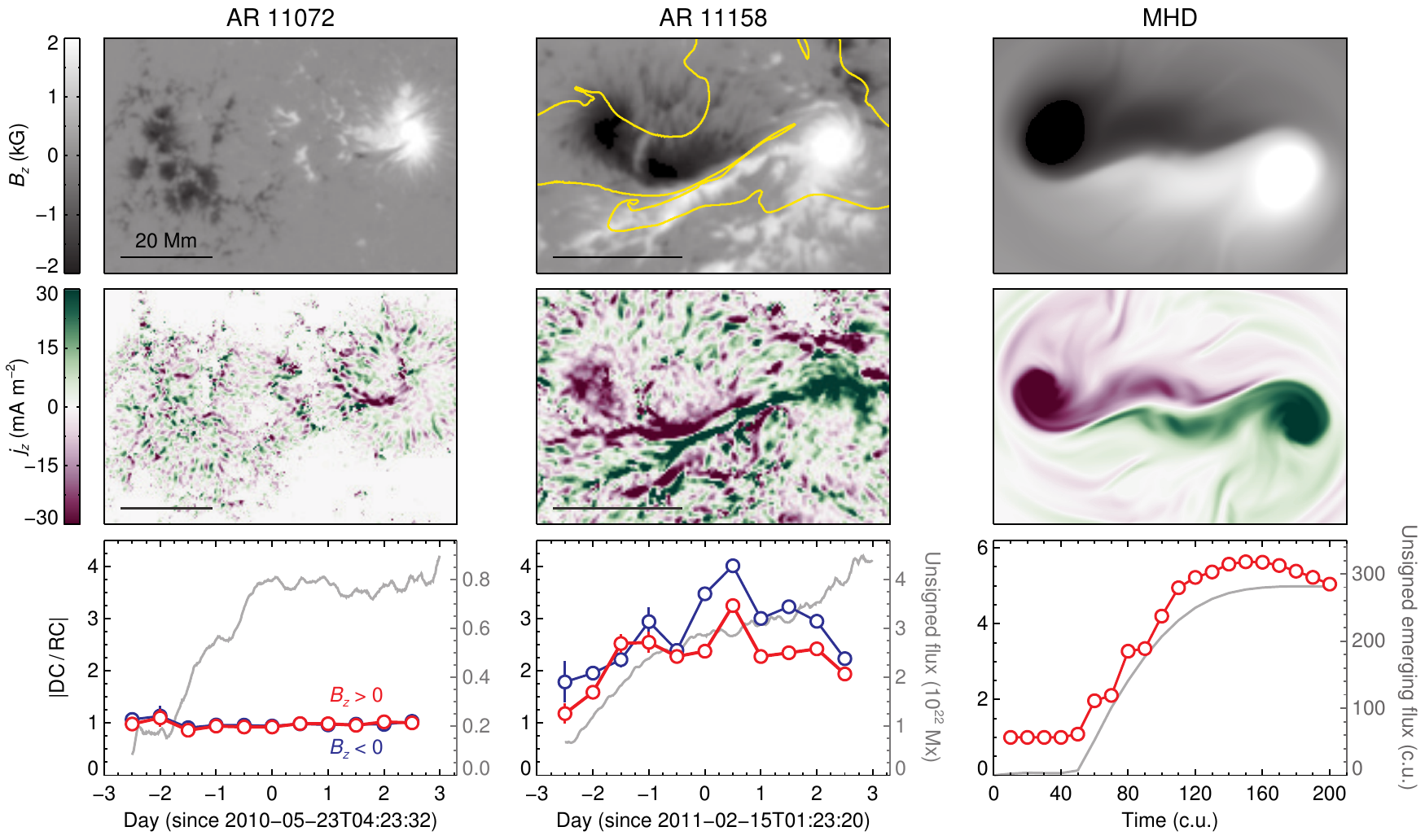}}
\caption
{
{ Left}, { middle:} $B_z$ and $J_z$ in quiet AR 11072 on 23 May 2010 and 
in the flare-productive center of AR 11158 on 15 February 2011. No core-region 
mask was used for the well-isolated, bipolar AR 11072. Bottom panels show the 
evolution of the unsigned magnetic flux (gray line) and the direct/return current 
ratio, $|DC/RC|$. Red (blue) curves show $|DC/RC|^+$ ($|DC/RC|^-$). Error bars 
are $3\sigma$, estimated from the inversion-uncertainty of the vector magnetic 
field data (errors smaller than circles are omitted).
{ Right:} Corresponding quantities for the MHD simulation by \cite{leake13}.
$|DC/RC|$ is equal for both polarities; only $|DC/RC|^+$ is plotted. Times and 
flux are in code units.  
}
\label{reversal}
\end{figure*}

\begin{figure*}
\centerline{\includegraphics{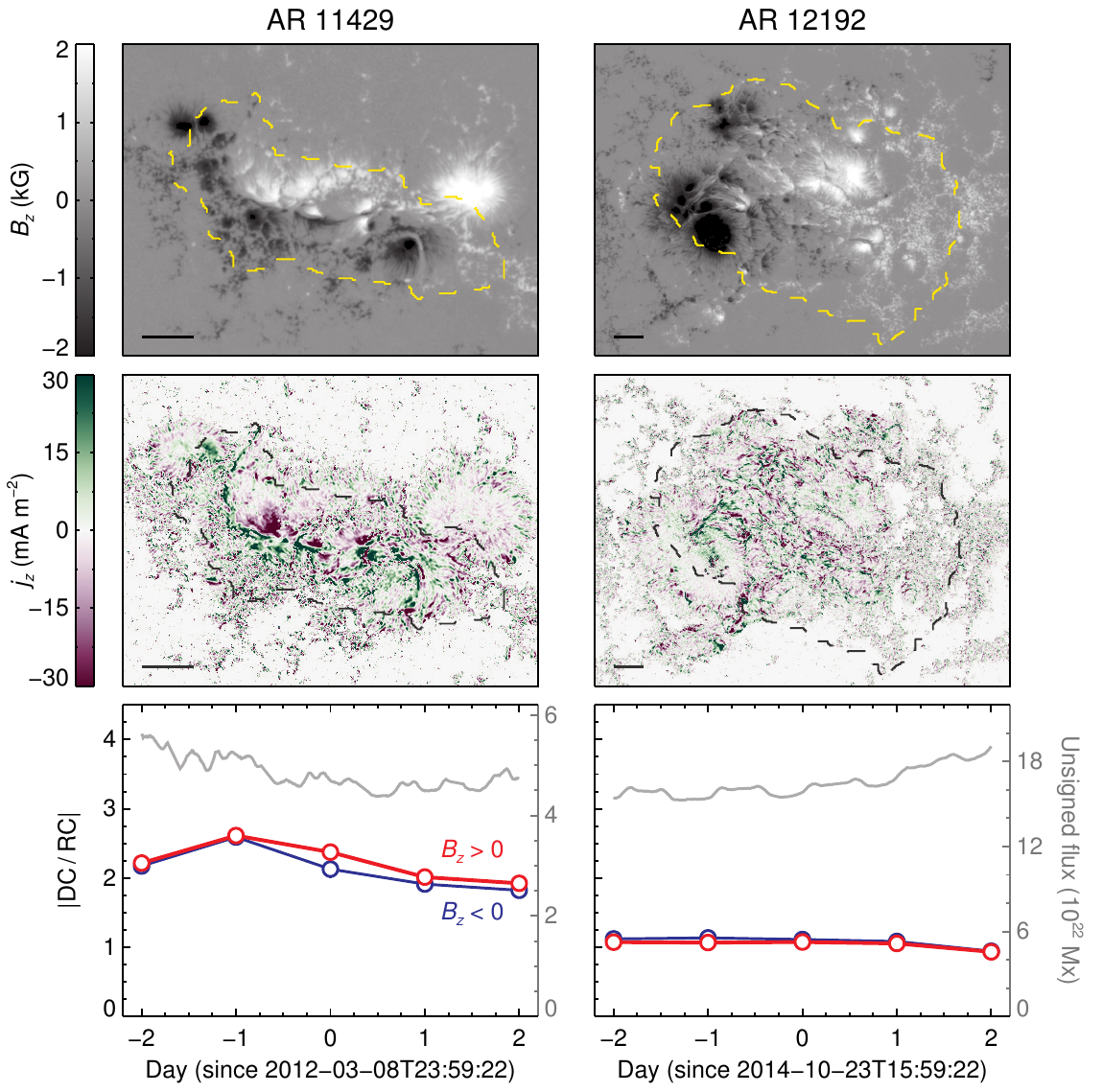}}
\caption
{
Same as Figure\,\ref{reversal} for the flare-productive centers of AR 11429 on 
9 March 2012 (left) and AR 12192 on 23 October 2014 (right). Dashed-line contours
enclose the core-region masks within which $|DC/RC|$ is computed.  
}
\label{twoars}
\end{figure*}

Our conclusions are quite speculative, of course, since our AR-sample is very small. Similar investigations using larger samples are required before more definite conclusions regarding the potential relationship between the amount of current-neutralization, PIL magnetic shear, and eruptive activity of solar active regions can be drawn. Furthermore, more numerical investigations are required, to gain a deeper physical understanding of the factors that determine the electric-current distribution in ARs and its role for eruptive activity. The very good agreement of the MHD simulation by \cite{leake13} with the observations of AR 11158 (Figure\,\ref{reversal}) demonstrates that such simulations, despite various idealizations, are well suited for this task.

\acknowledgments 
We thank the SDO team members for their hard work! This work was supported by NASA Contract NAS5-02139 (HMI) to Stanford University. Data were used by courtesy of NASA/SDO and the HMI science team. T.T. was supported by NASA's LWS program and NSF grant AGS-1348577, J.E.L. by NASA's LWS and HSR programs, and V.S.T. by NASA's HSR program and NSF grant AGS-1560411.


\begin{thebibliography}{39}
\expandafter\ifx\csname natexlab\endcsname\relax\def\natexlab#1{#1}\fi

\bibitem[{{Alissandrakis}(1981)}]{alissandrakis81}
{Alissandrakis}, C.~E. 1981, \aap, 100, 197

\bibitem[{Aschwanden {et~al.}(2014)Aschwanden, Sun, \& Liu}]{ash14}
Aschwanden, M.~J., Sun, X., \& Liu, Y. 2014, Astrophys. J. L., 785, article id.
  34

\bibitem[{{Cheng} \& {Ding}(2016)}]{cheng.x16}
{Cheng}, X., \& {Ding}, M.~D. 2016, \apjs, 225, 16

\bibitem[{{Dalmasse} {et~al.}(2015){Dalmasse}, {Aulanier}, {D{\'e}moulin},
  {Kliem}, {T{\"o}r{\"o}k}, \& {Pariat}}]{dalmasse15}
{Dalmasse}, K., {Aulanier}, G., {D{\'e}moulin}, P., {Kliem}, B.,
  {T{\"o}r{\"o}k}, T., \& {Pariat}, E. 2015, \apj, 810, 17

\bibitem[{{Forbes}(2010)}]{forbes10}
{Forbes}, T. 2010, in Heliophysics: Space Storms and Radiation: Causes and
  Effects, ed. C.~J. {Schrijver} \& G.~L. {Siscoe} (Cambridge (UK): Cambridge
  University Press), 159

\bibitem[{{Gary}(1989)}]{gary89}
{Gary}, G.~A. 1989, \apjs, 69, 323

\bibitem[{{Georgoulis} {et~al.}(2012){Georgoulis}, {Titov}, \&
  {Miki{\'c}}}]{georgoulis12}
{Georgoulis}, M.~K., {Titov}, V.~S., \& {Miki{\'c}}, Z. 2012, \apj, 761, 61

\bibitem[{{Gosain} {et~al.}(2014){Gosain}, {D{\'e}moulin}, \& {L{\'o}pez
  Fuentes}}]{gosain14}
{Gosain}, S., {D{\'e}moulin}, P., \& {L{\'o}pez Fuentes}, M. 2014, \apj, 793,
  15

\bibitem[{Hagyard {et~al.}(1984)Hagyard, Teuber, West, \& Smith}]{hag84}
Hagyard, M.~J., Teuber, D., West, E.~A., \& Smith, J.~B. 1984, \solphys, 91,
  115

\bibitem[{Hoeksema {et~al.}(2014)Hoeksema, Liu, Hayashi, Sun, Schou, Couvidat,
  Elliot, Barnes, \& Leka}]{hoe14}
Hoeksema, J.~T., Liu, Y., Hayashi, K., Sun, X., Schou, J., Couvidat, S.,
  Elliot, R.~C., Barnes, G., \& Leka, K.~D. 2014, Sol. Phys., 289, 3483

\bibitem[{{Janvier} {et~al.}(2014){Janvier}, {Aulanier}, {Bommier},
  {Schmieder}, {D{\'e}moulin}, \& {Pariat}}]{janvier14a}
{Janvier}, M., {Aulanier}, G., {Bommier}, V., {Schmieder}, B., {D{\'e}moulin},
  P., \& {Pariat}, E. 2014, \apj, 788, 60

\bibitem[{{Kliem} \& {T{\"o}r{\"o}k}(2006)}]{kliem06}
{Kliem}, B., \& {T{\"o}r{\"o}k}, T. 2006, \prl, 96, 255002

\bibitem[{{Leake} {et~al.}(2013){Leake}, {Linton}, \&
  {T{\"o}r{\"o}k}}]{leake13}
{Leake}, J.~E., {Linton}, M.~G., \& {T{\"o}r{\"o}k}, T. 2013, \apj, 778, 99

\bibitem[{Liu \& Schuck(2012)}]{liu12}
Liu, Y., \& Schuck, P.~W. 2012, Astrophys. J., 761, article id. 105

\bibitem[{Liu {et~al.}(2014)Liu, Richardson, Wang, \& J.~G.~Luhmann}]{liu14}
Liu, Y.~D., Richardson, J.~D., Wang, C., \& J.~G.~Luhmann, J.~G. 2014,
  Astrophys. J. L., 788, article id. L28

\bibitem[{{Longcope} \& {Welsch}(2000)}]{longcope00}
{Longcope}, D.~W., \& {Welsch}, B.~T. 2000, \apj, 545, 1089

\bibitem[{{Mackay} {et~al.}(2010){Mackay}, {Karpen}, {Ballester}, {Schmieder},
  \& {Aulanier}}]{mackay10}
{Mackay}, D.~H., {Karpen}, J.~T., {Ballester}, J.~L., {Schmieder}, B., \&
  {Aulanier}, G. 2010, \ssr, 151, 333

\bibitem[{Melrose(1991)}]{mel91}
Melrose, D.~B. 1991, Astrophys. J., 381, 306

\bibitem[{Melrose(1995)}]{mel95}
---. 1995, Astrophys. J., 451, 391

\bibitem[{Parker(1996)}]{par96b}
Parker, E.~N. 1996, Astrophys. J., 471, 485

\bibitem[{{Pesnell} {et~al.}(2012){Pesnell}, {Thompson}, \&
  {Chamberlin}}]{pesnell12}
{Pesnell}, W.~D., {Thompson}, B.~J., \& {Chamberlin}, P.~C. 2012, \solphys,
  275, 3

\bibitem[{{Petrie}(2013)}]{petrie13}
{Petrie}, G.~J.~D. 2013, \solphys, 287, 415

\bibitem[{{Ravindra} {et~al.}(2011){Ravindra}, {Venkatakrishnan}, {Tiwari}, \&
  {Bhattacharyya}}]{ravindra11}
{Ravindra}, B., {Venkatakrishnan}, P., {Tiwari}, S.~K., \& {Bhattacharyya}, R.
  2011, \apj, 740, 19

\bibitem[{Scherrer {et~al.}(2012)Scherrer, Schou, Bush, Kosovichev, Bogart,
  Hoeksema, Liu, Duvall, Zhao, Title, Schrijver, Tarbell, \& Tomczyk}]{sche12}
Scherrer, P.~H., Schou, J., Bush, R.~I., Kosovichev, A.~G., Bogart, R.~S.,
  Hoeksema, J.~T., Liu, Y., Duvall, Jr., T.~L., Zhao, J., Title, A.~M.,
  Schrijver, C.~J., Tarbell, T.~D., \& Tomczyk, S. 2012, Sol. Phys., 275, 207

\bibitem[{Schou {et~al.}(2012)Schou, Scherrer, Bush, Wachter, Couvidat,
  Rabello-Soares, Bogart, Hoeksema, Liu, Duvall, Akin, Allard, Miles, Rairden,
  Shine, Tarbell, Title, Wolfson, Elmore, Norton, \& Tomczyk}]{scho12}
Schou, J., Scherrer, P.~H., Bush, R.~I., Wachter, R., Couvidat, S.,
  Rabello-Soares, M.~C., Bogart, R.~S., Hoeksema, J.~T., Liu, Y., Duvall, Jr.,
  T.~L., Akin, D.~J., Allard, B.~A., Miles, J.~W., Rairden, R., Shine, R.~A.,
  Tarbell, T.~D., Title, A.~M., Wolfson, C.~J., Elmore, D.~F., Norton, A.~A.,
  \& Tomczyk, S. 2012, Sol. Phys., 275, 229

\bibitem[{Sun {et~al.}(2015)Sun, Bobra, Hoeksema, Liu, Li, Shen, Couvidat,
  Norton, \& Fisher}]{sun15}
Sun, X., Bobra, M.~G., Hoeksema, J.~T., Liu, Y., Li, Y., Shen, C., Couvidat,
  S., Norton, A.~A., \& Fisher, G.~H. 2015, Astrophys. J. L., 804, L28

\bibitem[{Sun {et~al.}(2012)Sun, Hoeksema, Liu, Wiegelmann, Hayashi, Chen, \&
  Thalmann}]{sun12}
Sun, X., Hoeksema, J.~T., Liu, Y., Wiegelmann, T., Hayashi, K., Chen, Q., \&
  Thalmann, J. 2012, Astrophys. J., 748, 77

\bibitem[{Thalmann {et~al.}(2015)Thalmann, Su, Temmer, \& Veronig}]{tha15}
Thalmann, J.~K., Su, Y., Temmer, M., \& Veronig, A.~M. 2015, Astrophys. J. L.,
  801, article id. L23

\bibitem[{Titov(2007)}]{tit07}
Titov, V.~S. 2007, Astrophys. J., 660, 863

\bibitem[{Titov {et~al.}(2011)Titov, Mikic, Linker, Lionello, \&
  Antiochos}]{tit11}
Titov, V.~S., Mikic, Z., Linker, J.~A., Lionello, R., \& Antiochos, S.~K. 2011,
  Astrophys. J., 731, 111

\bibitem[{{T{\"o}r{\"o}k} \& {Kliem}(2003)}]{torok03}
{T{\"o}r{\"o}k}, T., \& {Kliem}, B. 2003, \aap, 406, 1043

\bibitem[{{T{\"o}r{\"o}k} {et~al.}(2014){T{\"o}r{\"o}k}, {Leake}, {Titov},
  {Archontis}, {Miki{\'c}}, {Linton}, {Dalmasse}, {Aulanier}, \&
  {Kliem}}]{torok14a}
{T{\"o}r{\"o}k}, T., {Leake}, J.~E., {Titov}, V.~S., {Archontis}, V.,
  {Miki{\'c}}, Z., {Linton}, M.~G., {Dalmasse}, K., {Aulanier}, G., \& {Kliem},
  B. 2014, \apjl, 782, L10

\bibitem[{{Vemareddy} {et~al.}(2015){Vemareddy}, {Venkatakrishnan}, \&
  {Karthikreddy}}]{vemareddy15}
{Vemareddy}, P., {Venkatakrishnan}, P., \& {Karthikreddy}, S. 2015, Research in
  Astronomy and Astrophysics, 15, 1547

\bibitem[{{Venkatakrishnan} \& {Tiwari}(2009)}]{venkatakrishnan09}
{Venkatakrishnan}, P., \& {Tiwari}, S.~K. 2009, \apjl, 706, L114

\bibitem[{Wang {et~al.}(1994)Wang, M.~W.~Ewell, Zirin, \& Ai}]{wan94}
Wang, H., M.~W.~Ewell, J., Zirin, H., \& Ai, G. 1994, \apj, 424, 436

\bibitem[{Wang {et~al.}(2012)Wang, Liu, \& Wang}]{wan12}
Wang, S., Liu, C., \& Wang, H. 2012, Astrophys. J. L., 757, article id. L5

\bibitem[{{Wheatland} {et~al.}(2000){Wheatland}, {Sturrock}, \&
  {Roumeliotis}}]{wheatland00}
{Wheatland}, M.~S., {Sturrock}, P.~A., \& {Roumeliotis}, G. 2000, \apj, 540,
  1150

\bibitem[{{Wiegelmann}(2004)}]{wiegelmann04}
{Wiegelmann}, T. 2004, \solphys, 219, 87

\bibitem[{{Zakharov} \& {Shafranov}(1986)}]{Zakharov1986}
{Zakharov}, L.~E., \& {Shafranov}, V.~D. 1986, in Reviews of Plasma Physics,
  ed. M.~A. {Leontovich}, Vol.~11, 153

\end{thebibliography}


\clearpage

\clearpage

\begin{deluxetable*}{llllllll}
\tablecaption{Magnetic shear at and around the PIL and direct/return current ratio for all ARs during their disk passage. The top (bottom) two ARs are emerging (well-developed).
\label{tbl-1}}
\tablewidth{0pt}
\tablehead{
\colhead{AR} & \colhead{Quantity$^a$} & \colhead{Day 1} & \colhead{Day 2} & \colhead{Day 3} & \colhead{Day 4} & \colhead{Day 5$^b$} & \colhead{Average$^c$}\\
}
\startdata
AR 11072 & $\Phi$ & $37.6^\circ \pm 22.8^\circ$  & $26.5^\circ \pm 15.0^\circ$ & $15.2^\circ \pm 7.4^\circ$ & $18.7^\circ \pm 11.7^\circ$ & $27.6^\circ \pm 8.7^\circ$ & $25.1^\circ \pm 6.4^\circ$\\
AR 11158 & $\Phi$ & $56.7^\circ \pm 16.5^\circ$  & $70.1^\circ \pm 10.0^\circ$ & $70.1^\circ \pm 13.8^\circ$ & $66.4^\circ \pm 16.5^\circ$ & $54.8^\circ \pm 18.8^\circ$ & $63.6^\circ \pm 6.9^\circ$\\
AR 11072 & $|DC/RC|$ & $1.0\pm 0.02$  & $0.95\pm 0.01$ & $0.96\pm 0.01$ & $0.97\pm 0.01$ & $1.01\pm 0.01$ & $0.98\pm 0.01$ \\
AR 11158 & $|DC/RC|$ & $1.63\pm 0.04$  & $2.55\pm 0.04$ & $2.63\pm 0.03$ & $3.13\pm 0.03$ & $2.74\pm 0.03$ & $2.54\pm 0.02$ \\
\hline
AR 11429 & $\Phi$ & $65.6^\circ \pm 23.3^\circ $  & $70.5^\circ \pm 17.3^\circ$ & $66.1^\circ \pm 17.7^\circ$ & $67.3^\circ \pm 18.3^\circ$  & $67.8^\circ \pm 17.8^\circ$ & $67.4^\circ \pm 8.5^\circ$\\
AR 12192 & $\Phi$ & $33.4^\circ \pm 23.9^\circ$  & $41.2^\circ \pm 26.3^\circ$ & $41.6^\circ \pm 25.0^\circ $ & $44.6^\circ \pm 25.6^\circ$ & $45.0^\circ \pm 26.2^\circ$ & $41.2^\circ \pm 11.3^\circ$\\
AR 11429 & $|DC/RC|$ & $2.19\pm 0.03$  & $2.60\pm 0.03$ & $2.23\pm 0.03$ & $1.96\pm 0.03$ & $1.87\pm 0.02$ & $2.17\pm 0.01$ \\
AR 12192 & $|DC/RC|$ & $1.09\pm 0.01$  & $1.10\pm 0.01$ & $1.09\pm 0.01$ & $1.07\pm 0.01$ & $0.94\pm 0.01$ & $1.06\pm 0.01$ \\
\enddata
\tablenotetext{a}{Computed quantities. $\Phi$ is the average magnetic shear angle at and around the PIL; $|DC/RC|$ the ratio of total direct and return current, where $|DC/RC|=(|DC/RC|^{+}+|DC/RC|^{-})/2$. For the two emerging ARs, the daily values are averaged over two data sets per day. 
}

\tablenotetext{b}{Day of a 5-day period when the ARs were visible on the solar disk. The time periods are 21--25 May 2010, 13--17 February 2011, 7--11 March 2012, and 21--25 October 2014 for ARs 11072, 11158, 11429, and 12192, respectively.}
\tablenotetext{c}{Values averaged over the 5-day period.}
\label{table1}
\end{deluxetable*}


\begin{deluxetable*}{llllll}
\tablecaption{Relevant AR properties. $\langle\,|DC/RC|\,\rangle$ and $\langle\,\Phi\,\rangle$ are 5-day average values; ``$J$-pattern'' refers to a coherent direct-current channel around the PIL (see text); ``Flares'' refers to strong flares of at least M-class.   
\label{tbl-2}}
\tablewidth{0pt}
\tablehead{\colhead{AR} & \colhead{$\langle\,|DC/RC|\,\rangle$} & \colhead{$\langle\,\Phi\,\rangle$} & $J$-pattern & \colhead{Flares} & \colhead{CMEs} \\}
\startdata
AR 11072  &  $0.98\pm 0.01$  &  $25.1^\circ \pm  6.4^\circ$  &  No   & No   &  No  \\
AR 12192  &  $1.06\pm 0.01$  &  $41.2^\circ \pm 11.3^\circ$  &  No   & Yes  &  No  \\
AR 11429  &  $2.17\pm 0.01$  &  $67.4^\circ \pm  8.5^\circ$  &  Yes  & Yes  &  Yes \\
AR 11158  &  $2.54\pm 0.02$  &  $63.6^\circ \pm  6.9^\circ$  &  Yes  & Yes  &  Yes \\
\enddata
\end{deluxetable*}

\end{document}